\documentclass[a4paper,11pt]{article}
\pdfoutput=1 

\usepackage{jcappub} 
                                        
\usepackage{aas_macros}
\usepackage{subcaption}
\usepackage[T1]{fontenc} 
\usepackage{float}
\usepackage[normalem]{ulem}
\usepackage{bm}
\usepackage{mathtools}
\usepackage{hyperref}

\newcommand{\avg}[1]{\ensuremath{\left\langle \,#1\, \right\rangle}}

\newcommand{\Mh}{\ensuremath{h^{-1}M_{\odot}}}

\newcommand{\Mpch}{\ensuremath{h^{-1}{\rm Mpc}} }
\newcommand{\hMpc}{\ensuremath{h\,{\rm Mpc}^{-1}}}

\newcommand{\fcoll}{\ensuremath{f_{\text{coll}}}}

\newcommand{\mhmin}{\ensuremath{M_{h, \text{ min}}}}
\newcommand{\qmhii}{\ensuremath{Q_\text{HII}^M}}
\newcommand{\Fcoll}{\ensuremath{\mathbf{f}_\text{coll}}}

\newcommand{\be}{\begin{equation}}
\newcommand{\ee}{\end{equation}}

\title{\boldmath Conditioning halos on the tidal environment for fast and accurate HI power spectra during reionization}

\usepackage{natbib}
\author[a, 1]{Gaurav Pundir,\note{Corresponding author.}}
\author[b]{Tirthankar Roy Choudhury}
\author[c]{and Aseem Paranjape}

\affiliation[a]{Department of Physics, \\ 
Indian Institute of Science Education and Research Pune, \\
Dr. Homi Bhabha Road, Pashan, Pune 411008, India}
\affiliation[b]{
National Centre for Radio Astrophysics, TIFR, \\
Post Bag 3, Ganeshkhind, Pune 411007, India}
\affiliation[c]{
Inter-University Centre for Astronomy \& Astrophysics, \\
Post Bag 4, Ganeshkhind, Pune 411007, India}

\emailAdd{gauravpundir909@gmail.com}
\emailAdd{tirth@ncra.tifr.res.in}
\emailAdd{aseem@iucaa.in}

\abstract{Predicting the statistical properties of the neutral hydrogen (HI) density field during reionization is an important step in using upcoming 21 cm observations to constrain models of reionization. Semi-numerical models of reionization are often coupled with the collapse fraction field $\fcoll(\mathbf{x})$, which determines the fraction of dark matter within halos. In this work, we improve upon earlier prescriptions that compute \fcoll\ based on the dark matter overdensity $\delta(\mathbf{x})$ alone, to include more information about the environment in the form of eigenvalues of the tidal tensor. We compute the mean of the \fcoll\ conditioned on these eigenvalues from a set of high-resolution, small-volume simulations and use them to sample the \fcoll\ field of a low-resolution, large-volume simulation. We subsequently use a semi-numerical code for reionization to compute the HI density field and its power spectrum, and benchmark our results against a reference high-resolution, large-volume simulation. Across variations in redshift, ionized fraction, grid resolution, and minimum halo mass, our method recovers the large-scale HI power spectrum with errors at the $\lesssim 2$\%--5\% level for $k \lesssim 0.5\ \hMpc$, providing a substantial improvement over the $\sim 10\%$ results previously obtained using density-only conditioning. Overall, this makes our method a simple yet efficient tool for forward modeling HI maps during reionization.}

\begin{document}
\maketitle
\flushbottom

\section{Introduction}
\label{sec:intro}

The Epoch of Reionization (EoR) brought about the end of the universe's `dark ages' and transformed it into the ionized, luminous expanse that we observe today. This era is a key frontier in modern cosmology, due to its links with the formation of the first luminous objects and subsequent structure formation \cite{gnedin_22_R, choudhury_22_R}. A primary observational probe for this epoch is the redshifted 21 cm hyperfine transition of neutral hydrogen (HI), which offers insight into the distribution of the HI gas in the intergalactic medium (IGM) during this period \cite{furlanetto_06_R, pritchard_12_R, mesinger_19_R}. The statistical properties of the 21 cm signal, particularly its power spectrum, contain information about cosmological and astrophysical parameters, making accurate theoretical models essential for interpreting the upcoming observational data. 

The most comprehensive theoretical approach to modeling the EoR involves running computationally intensive radiative transfer (RT) simulations that explicitly track the complex interactions between matter and ionizing photons \cite{gnedin_00, ciardi_03, mellema_06, trac&cen_07, gnedin_14, rosdahl_18, lewis_22}. However, these simulations face a huge computational memory challenge due to the need for a high-dynamic range --- they must simultaneously resolve the smallest luminous sources (corresponding to dark matter halos of mass $\sim 10^8\ \Mh$) and cover a large enough volume to statistically sample the distribution of ionized bubbles \cite{iliev_14, kaur_20}.  

As a result, in order to limit the computational expense and make parameter space exploration feasible, semi-numerical models of reionization that are faster and bypass the full physics of radiative transfer have been developed. These models often rely on an excursion-set approach \cite{BCEK_91} and a photon-counting argument to predict the ionization field \cite{FZH_04, mesinger_07, zahn_07, choudhury_09, mesinger_11, lin_16, choudhury_18}. When coupled with dark-matter-only N-body simulations, these models require an input known as the collapse fraction field $\fcoll(\mathbf{x})$, which quantifies the fraction of dark matter residing in halos at each location. While semi-analytical prescriptions, such as the conditional Press-Schechter \cite{PS_74, BCEK_91} and Sheth-Tormen \cite{ST_99, ST_02} mass functions, can be used to generate the \fcoll\ field, they are known to be approximations of the more complex physics of halo formation and are quite inaccurate compared to N-body simulations at the redshifts relevant to reionization \cite{reed_07, tinker_08, courtin_10, crocce_11}. 

On the other hand, computing the collapse fraction field directly from large-volume and high-resolution N-body simulations, to be input into semi-numerical codes of reionization, runs into the same dynamic range problem as mentioned before. Therefore, there have been attempts to combine low-dynamic range simulations (with a lower computational cost) in a way that uses the large volume of a low-resolution simulation and the properly resolved halos of a low-volume, high-resolution simulation \cite{ahn_12, iliev_14}. This has also been the methodology of our previous work \cite{pundir_25} (henceforth Paper1), which had the same goal as the current one and where we also incorporated stochasticity in the collapse fraction predictions. 

However, all of these works compute the collapse fraction by taking the local dark matter density contrast $\delta$ alone as a proxy for the cosmological environment. The formation and clustering of halos are not just functions of density but are significantly influenced by anisotropic gravitational forces at large scales. Interesting alternatives that utilize information beyond the matter density field include constructing the tidal tensor and classifying different cosmic environments by comparing its eigenvalues with a threshold, first proposed in \cite{hahn_07} and further explored in \cite{forero-romero_09, bonnaire_22}. Such a classification motivates the use of the tidal tensor eigenvalues to provide a more environmentally-informed prediction of \fcoll, which could in turn produce a more accurate HI density map. An instance of this approach can be found in \cite{barsode_24}, where the authors populate a low-resolution, large-volume box with low-mass halos taken from a small-volume, high-resolution simulation by `matching' the cells based on their tidal tensor eigenvalues. This has the disadvantage of requiring both the large and small boxes simultaneously to make the full \fcoll\ prediction. 

In this work, we set out with the same goal as that of Paper1 --- to accurately and efficiently forward model the HI power spectrum during reionization. We use a similar approach of combining low dynamic range simulations to produce a high-fidelity $\fcoll(\mathbf{x})$ field, to be used as an input for the semi-numerical code for reionization \textsc{script} to get the HI density map. The crucial difference, however, is that we now condition the \fcoll\ values on linear combinations of the three eigenvalues of the tidal tensor.\footnote{The \fcoll\ distribution is assumed to correlate only with the eigenvalues $\{\lambda_1,\lambda_2,\lambda_3\}$ of the local tidal tensor, an approximation we justify later in the text. Throughout, we will be interested in the mean value of the conditional distribution $p(\fcoll|\lambda_1,\lambda_2,\lambda_3)$, which is what we colloquially refer to as the (mean of) \fcoll\ values conditioned on the tidal tensor.} For brevity, we henceforth refer to these mean \fcoll\ values as the \textit{conditional means}, highlighting their dependence on the eigenvalues while noting that they are just the expectation values of the full conditional distribution. We thus focus on a deterministic sampling method that ignores the scatter in \fcoll\ for a given set of eigenvalues. This results in a very simple method that does not involve any complex machine learning algorithm, is computationally efficient, and still ends up producing substantially better results than the GPR-based method of Paper1 for the large-scale HI power spectrum. 

The paper has been organized as follows --- section \ref{sec:sims} describes all the simulations used as well as the quantities that are defined within them, section \ref{sec:method} provides some motivation along with the details and optimization schema of the algorithm, section \ref{sec:results} presents the results for the HI map and the HI power spectrum across a range of parameters, section \ref{sec:disc} discusses and compares the results with the previous work, and section \ref{sec:conc} concludes the paper. Appendix \ref{app:robustness} provides a check on the sensitivity of the results to the binning parameters.

\section{Simulations}
\label{sec:sims}

We run three different kinds of N-body cosmological simulations for the purposes of storing the conditional mean \fcoll\ values, referencing them to make a prediction, and testing the accuracy of the prediction. All the simulations have been run using the GADGET-2\footnote{\href{https://wwwmpa.mpa-garching.mpg.de/galform/gadget/}{https://wwwmpa.mpa-garching.mpg.de/galform/gadget/}} \cite{springel_05} code. The cosmological parameters used are $\Omega_m=0.308, H_0 = 67.8 \text{ km s}^{-1} \text{ Mpc}^{-1}, \sigma_8 = 0.829, \text{ and } n_s = 0.961$ in a flat, $\Lambda$CDM cosmology following the results from Planck \cite{planck_18}. We locate the positions and masses of the dark matter halos using a Friends-of-Friends (FoF) halo finder \cite{davis_85}. We define a grid with a length-scale $\Delta x$ over the simulation boxes, and for each grid cell compute the dark matter overdensity $\delta(\mathbf{x})$ using a cloud-in-cell (CIC) mass assignment scheme and the collapse fraction field denoted by \fcoll\ in the same way as in Paper1, 
\begin{equation}\label{fcoll_eqn}
    \fcoll (\mathbf{x}) = \dfrac{\sum_h m_h (\mathbf{x})}{M_{\text{tot}}(\mathbf{x})}\,,
\end{equation}
with the sum carried over all the halos inside the cell under consideration that are above the minimum halo mass cutoff \mhmin. Additionally, this time we also incorporate information contained in the tidal field, which is given at each point by the Hessian of the Newtonian gravitational potential $\Phi(\mathbf{x})$ as
\begin{equation}\label{tidal_tensor_eqn}
    T_{ij} = \dfrac{\partial^2\Phi}{\partial x_i \partial x_j}~ (i, j \in \{1, 2, 3\}) \,,
\end{equation}
where $x_i$ denotes the $i^{\text{th}}$ Cartesian component of the position vector $\mathbf{x}$. Specifically, we are interested in the three eigenvalues of the tidal tensor $T_{ij}$ denoted by $\lambda_1, \lambda_2, \lambda_3$, ordered such that $\lambda_1 \leq \lambda_2 \leq \lambda_3$. From Poisson's equation, and using the fact that the sum of eigenvalues of a matrix is equal to its trace, we have
\begin{equation}\label{poisson_eqn}
    \nabla^2 \Phi = \delta = \lambda_1 + \lambda_2 + \lambda_3 \,,
\end{equation}
where the potential has been appropriately scaled by $4\pi G \Bar{\rho}$. Using the Fourier transform of the CIC overdensity field $\delta(\mathbf{k})$, we can solve Poisson's equation and substitute $\Phi(\mathbf{k})$ in equation \ref{tidal_tensor_eqn} to compute the tidal tensor in Fourier space, and after transforming it back to position space, the three eigenvalues at each cell. Instead of working with the eigenvalues as is, we choose to define the following linear combinations that are all strictly non-negative --
\begin{align}
    \ell_1 &= 1 + \lambda_1 + \lambda_2 + \lambda_3 = 1 + \delta \label{lin_combs_eqn1} \\
    \ell_2 &= \lambda_2 - \lambda_1 \label{lin_combs_eqn2} \\
    \ell_3 &= \lambda_3 - \lambda_2 \label{lin_combs_eqn3}
\end{align}

Now let us outline the details of various simulation boxes -- 

\begin{enumerate}
    \item \textbf{Small Boxes (SB):} These are supposed to be small-volume but high-resolution, and are run with a number of particles $N=512^3$ and a volume $V = (40\ \Mpch)^3$. We run seven of these, which are collectively used to obtain the conditional means of \fcoll\ conditioned on the three variables $\ell_1, \ell_2, \ell_3$, denoted by \avg{\fcoll | \bm{\ell}}. Running a single such box takes $\sim 210$ CPU hours and a maximum 20 GB of RAM. 
    \item \textbf{Large Box (LB):} This is supposed to be large-volume but low-resolution, and is run with a number of particles $N = 512^3$ and a volume $V = (80\ \Mpch)^3$. We run a single such box and the $\bm{\ell} = (\ell_1, \ell_2, \ell_3)$ values for each cell will be used to assign a predicted \fcoll\ to that cell, using the conditional mean computed from the SB. Running this takes $\sim 220$ CPU hours and a maximum 20 GB of RAM. 
    \item \textbf{Reference Box (RB):} This is a single large-volume and high-resolution box, run with a number of particles $N = 1024^3$ and a volume $V = (80\ \Mpch)^3$. We compute the \fcoll\ field of the RB, which is a higher dynamic range simulation, to benchmark the accuracy of our prediction made by combining information from LB and SB. Running this takes $\sim 2900$ CPU hours and a maximum 160 GB of RAM. Note that our attempt of using SB and LB combined requires less RAM than running a single RB. Once the conditional means from the SB are available, one has to only run the LB. 
\end{enumerate}

\section{Methodology}
\label{sec:method}

Previous studies attempting to classify different structures in the cosmic web \cite{hahn_07, forero-romero_09} note that a dynamically motivated algorithm based on the Hessian of the gravitational potential (the tidal tensor defined in equation \ref{tidal_tensor_eqn} above) serves a better purpose than a purely geometrical classification based on the density field alone. This is related to the fact that the linearized equation of motion for a test particle near a gravitational potential extremum will be dictated by the tidal field eigenvalues \cite{hahn_07}. Since masses of dark matter halos correlate with the local environment \cite{lemson_99}, one can expect that a good proxy of the cosmic environment would also contain information regarding the total halo mass at each location, thereby motivating the use of the tidal tensor eigenvalues to compute the collapse fraction \fcoll\ at each location.

\subsection{Binning and Computing the Conditional Means}

In order to bin the three positive variables $\ell_1, \ell_2, \ell_3$, we face a similar issue as described in subsection 3.1 of Paper1, which is that the distribution is highly skewed with a long tail towards higher values. Therefore, we adopt a similar procedure of binning in logspace, i.e. over the variables $\log(\ell_1), \log(\ell_2), \log(\ell_3)$. For simplicity, we assume uniform binning in logspace this time instead of a variable binning where the bin width increases away from the centre. 
                 
We postpone the discussion of how to choose the number of bins for each variable to subsection \ref{subsec:optim}. For now, assume that the binning scheme has been fixed and the number of bins along the three variables are $N_1, N_2$ and $N_3$. Thus, from the binning we have an $N_1 \times N_2 \times N_3$ matrix where each cell represents the three-dimensional bin in the space spanned by $\log(\bm{\ell}) \coloneq (\log(\ell_1), \log(\ell_2), \log(\ell_3))$. Let us identify each such 3d bin by its `bin-centre', which is simply the 3-tuple of the bin-centres of the logarithmic variable along each direction, $(\log(\ell_{1m}), \log(\ell_{2m}), \log(\ell_{3m}))$ or more simply, $\log(\bm{\ell}_m)$, where $m$ can range from 1 upto $N_\alpha$ for the variable $\log(\ell_\alpha)$. For each such 3d bin, we retrieve indices of all the $\log(\bm{\ell})$ values combined over the seven realizations of SB that belong to that bin. We then use the same indexing on the corresponding list of combined \fcoll\ values from SB and compute their mean. This is the conditional mean to be assigned to the respective bin, and can be denoted by \avg{\fcoll| \bm{\ell}_m} for that bin. Once the process is done for each bin, we have the three-dimensional \textit{conditional mean matrix}, \avg{\Fcoll | \bm{\ell}} of size $N_1 \times N_2 \times N_3$. 

\subsection{Sampling}

Given that \avg{\fcoll | \bm{\ell}_m} has been computed and stored for each bin, we can use the $\bm{\ell}$ from each cell of the LB as an input to make its corresponding \fcoll\ prediction. This amounts to the assumptions that the local \fcoll\ distribution depends entirely on the local $\bm{\ell}$ values, and that the seven SB simulations provide a statistically robust computation of the conditional means. For each cell in LB, we simply find the index in the conditional mean matrix of the bin where its $\bm{\ell}$ values lie. If the centre of this bin is $\bm{\ell}_0$, the corresponding \fcoll\ from the matrix, \avg{\fcoll | \bm{\ell}_0}, is assigned as the prediction. This way, we generate a full three-dimensional \textit{predicted} $\fcoll(\mathbf{x})$ field. It is worth emphasizing the computational simplicity of this process, where a combination of some optimized binning and indexing allows us to construct both the conditional mean matrix and the full LB prediction within just a few minutes. 

\subsection{Optimization of Binning} \label{subsec:optim}

A binning scheme consisting of too few bins may wash out crucial environmental information by over-smoothing the conditional means, while too many can lead to statistical noise if individual bins are sparsely populated. To balance this, it is necessary to optimize the number of bins along each axis to minimize the error in the final HI power spectrum. The binning scheme we use is uniform in the three variables $(\log(\ell_1), \log(\ell_2), \log(\ell_3))$ and is thus decided solely by the number of bins along each direction. We input the predicted \fcoll\ field corresponding to a particular binning scheme into the semi-numerical code for reionization called \textsc{script}, whose details are described in section \ref{sec:results}, to get the neutral hydrogen (or HI) density field. The metric for the binning scheme would be the accuracy of the power spectrum $P_\text{HI}(k)$ of this HI density field, or $\rho_\text{HI}(\mathbf{x})$, defined via
\begin{equation}\label{Pk_eqn}
\frac{\avg{\rho_\text{HI}(\mathbf{k})\rho_\text{HI}^*(\mathbf{k}')}}{\overline{\rho_\text{HI}}^2} = (2\pi)^3P_\text{HI}(k)\delta_D(\mathbf{k}-\mathbf{k}') \,,
\end{equation}
where $\rho_\text{HI}(\mathbf{k})$ denotes the Fourier conjugate of the density field, an asterisk denotes complex conjugation, $\delta_D$ is the Dirac delta function, angular brackets represent averaging in Fourier space, and $\overline{\rho_\text{HI}}$ is the mean of the HI density field in position space.  

The accuracy of $P_\text{HI}(k)$ computed this way is to be checked against that of the \textit{true} HI density field, which is obtained by putting the \fcoll\ field from the RB (section \ref{sec:sims}; henceforth the `true' \fcoll\ field) into \textsc{script}. The optimization criterion consists of finding which binning scheme produces the least relative error at large scales, or low $k$ (the reason to focus on large scales will be clear in section \ref{sec:results}). Further details of this procedure and the final, optimized binning schemes are discussed in the following section and appendix \ref{app:robustness}.

\section{Results}
\label{sec:results}

We employ the \textbf{S}emi-numerical \textbf{C}ode for \textbf{R}e\textbf{I}onization with \textbf{P}ho\textbf{T}on-conservation (\textsc{script})\footnote{\href{https://bitbucket.org/rctirthankar/script
}{https://bitbucket.org/rctirthankar/script
}} \cite{choudhury_18} to generate the neutral hydrogen density fields from the predicted or true collapse fraction fields. The code takes as an input the reionization efficiency parameter $\zeta$ (apart from the collapse fraction field $\fcoll(\mathbf{x})$) which represents the number of ionizing photons entering the intergalactic medium per hydrogen atom in dark matter halos. \textsc{script} models the process of reionization by constructing ionized (HII) bubbles around sources of ionizing radiation. A key feature is its explicit enforcement of photon conservation; it allows regions to be ‘over-ionized’ initially, and then redistributes these excess photons to neighboring neutral regions. This process is iterated until the ionization fraction $x_\text{HII}(\mathbf{x})$ in all cells is less than or equal to unity, ensuring that the resulting large-scale statistics of
the ionization field are robust to the grid resolution $\Delta x$.

The output of \textsc{script} is the ionization fraction field $x_{\text{HII}}(\mathbf{x})$, from which we derive the neutral hydrogen fraction field as $x_{\text{HI}}(\mathbf{x}) = 1 - x_{\text{HII}}(\mathbf{x})$. We then compute the mass-averaged neutral hydrogen density field as
\begin{equation}
x_{\text{HI}}^M(\mathbf{x}) = x_{\text{HI}}(\mathbf{x})(1 + \delta(\mathbf{x})) \propto \rho_{\text{HI}}(\mathbf{x}) \,,
\end{equation}
where $\delta(\mathbf{x})$ is the matter density contrast. The mass-averaged ionized hydrogen density field $x_{\text{HII}}^M(\mathbf{x})$ can be computed analogously, and the global ionization fraction is defined as $\qmhii \equiv \langle x_{\text{HII}}^M(\mathbf{x}) \rangle$, where angular brackets denote a spatial average over the whole box. \qmhii\ is related to the global neutral fraction $\langle x_{\text{HI}}^M(\mathbf{x}) \rangle$ as $\langle x_{\text{HI}}^M(\mathbf{x}) \rangle = 1 - \qmhii$. By applying this procedure to the \textit{predicted} \fcoll\ field from our method and the \textit{true} \fcoll\ field from the RB, we can visually compare the corresponding HI density maps, or $x_{\text{HI}}^M(\mathbf{x})$, and also substitute them in equation \ref{Pk_eqn} to compute the true and predicted HI power spectra. We do this while adjusting the $\zeta$ for both the true and predicted fields such that they have the same ionization fraction \qmhii. We declare the fiducial case to be the same as in Paper1, with a redshift $z=7$, ionization fraction $\qmhii = 0.5$, grid scale $\Delta x = 0.5\ \Mpch$, and minimum halo mass $M_{h, \text{ min}} = 4.08 \times 10^8\ \Mh$. 

To assess the robustness and applicability of our method across the parameter space relevant to reionization studies, we examine the following variations in each of these parameters while keeping the others fixed at their fiducial values --
\begin{itemize}
    \item \textbf{Redshift:} varied from 7 (fiducial) to 5 and 9, covering the range from late to early reionization.
    \item \textbf{Ionized fraction:} varied from 0.5 (fiducial) to 0.25 and 0.75, to check for alternate reionization histories.
    \item \textbf{Grid size:} varied from 0.5 (fiducial) to 0.25 and 1 (in \Mpch), to examine the sensitivity to spatial resolution.
    \item \textbf{Minimum halo mass:} varied from 4.08 (fiducial) to 16.3 and 32.6 (in $10^8\ \Mh$), corresponding to different assumptions about the efficiency of star formation in low-mass halos. This is done by changing the minimum number of particles contained in a halo in the FoF halo finder from 10 (fiducial) to 40 and 80.
\end{itemize}

We wish to find an optimal binning scheme for each of these variations. For this, we compute the HI power spectra error as described in subsection \ref{subsec:optim} for every binning scheme defined by the number of bins $(n_1, n_2, n_3)$ along the three logarithmic variables, where $n_i$ is picked from \{10, 15, 20, 25, 30\}, independently for $i= 1, 2, 3$. This gives us a total of 125 schemes starting from $(10, 10, 10), (10, 10, 15) \dots$ till $(30, 30, 25), (30, 30, 30)$. For the $z$ variation, say, we select the scheme which shows a consistently low error in $P_\text{HI}(k)$ across all the cases $z=5, 7,$ and 9. This scheme may not be the one that gives the lowest error for each of these redshift cases separately, but for simplicity we choose the same scheme for all the cases of a given parameter variation, and the difference is insubstantial. Optimal schemes are chosen similarly for each of the other three parameter variations. 

Through this extensive process, we identify two distinct binning schemes that perform \textit{optimally} for different parameter variations. For variations in redshift $(z)$ and minimum halo mass $(M_{h, \text{ min}})$, the neutral hydrogen power spectra achieve excellent accuracy with a binning configuration of $(20, 15, 30)$ bins in $(\log(\ell_1), \log(\ell_2), \log(\ell_3))$ respectively, which we designate as \textit{binning scheme A}. Conversely, for variations in the global ionization fraction (\qmhii) and grid resolution $(\Delta x)$, optimal performance is obtained with $(25, 15, 20)$ bins, referred to as \textit{binning scheme B}. We check the robustness of our results to small variations around these two optimal schemes in appendix \ref{app:robustness}.

\begin{figure}
\centering
\includegraphics[width=\linewidth]{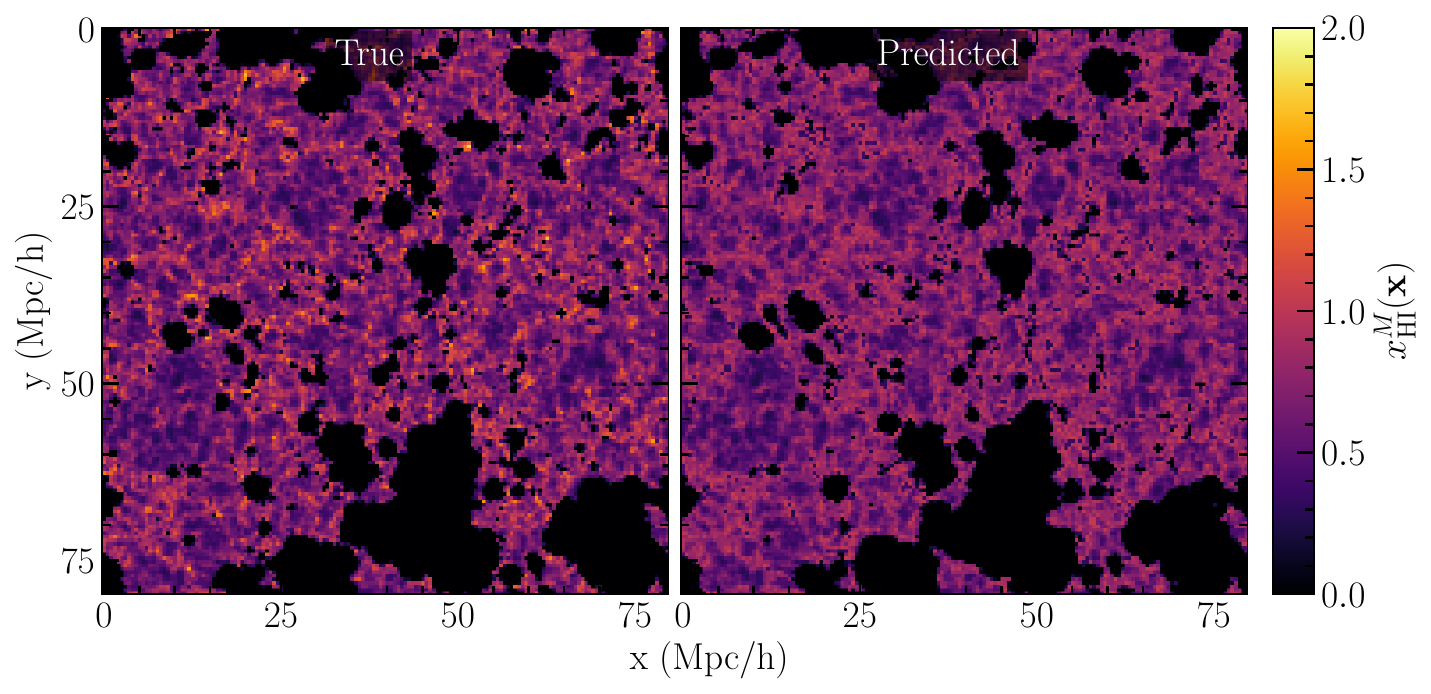}
\caption{\small The neutral hydrogen density field $x_{\text{HI}}^M(\mathbf{x})$ at $\qmhii = 0.5$ for both the truth (\textit{left panel}) and prediction (\textit{right panel}), shown for a slice through $z = 50\ \Mpch$ in our simulation volume. The maps are produced for the fiducial case, $z=7, \qmhii = 0.5,\Delta x = 0.5\ \Mpch, M_{h, \text{ min}} = 4.08 \times 10^8\ \Mh$. The black regions correspond to ionized bubbles where $x_{\text{HI}} \approx 0$. The maps demonstrate the inhomogeneous topology of reionization, with ionized regions preferentially forming around high-density regions that host the sources of ionizing photons. The predicted HI map lacks a lot of small-scale features as a direct consequence of our deterministic sampling method of using the conditional mean \fcoll\ values while averaging out its spread.}
\label{fig:hi_maps}
\end{figure}

We can now compare a 2d slice of the HI map between truth and prediction, generated for the fiducial case of parameters using binning scheme A. This is shown in figure \ref{fig:hi_maps}. While the large-scale structure of the HI density field matches quite well, one can notice a discrepancy at small-scales, where the predicted field seems to be a lot smoother than the true field. This is not surprising given that our sampling method for the \fcoll\ was based on a single, conditional mean value computed for a fixed $(\ell_1, \ell_2, \ell_3)$. This implies that any possible spread in the \fcoll\ due to variations in environment not captured by the tidal tensor eigenvalues at the grid scale was averaged out, producing a smoother \fcoll\ map with less variations than in the truth at small-scales, and correspondingly a smoother HI density map. This simply does not affect the large-scale topology as much because in bigger regions on average, the full distribution itself converges to the mean value. This issue is the same as that encountered in the \textit{deterministic} case of Paper1, described in detail in its Discussion section. 

\begin{figure}[h]
  \centering
  \begin{subfigure}{0.49\linewidth}
    \centering
    \includegraphics[width=\linewidth]{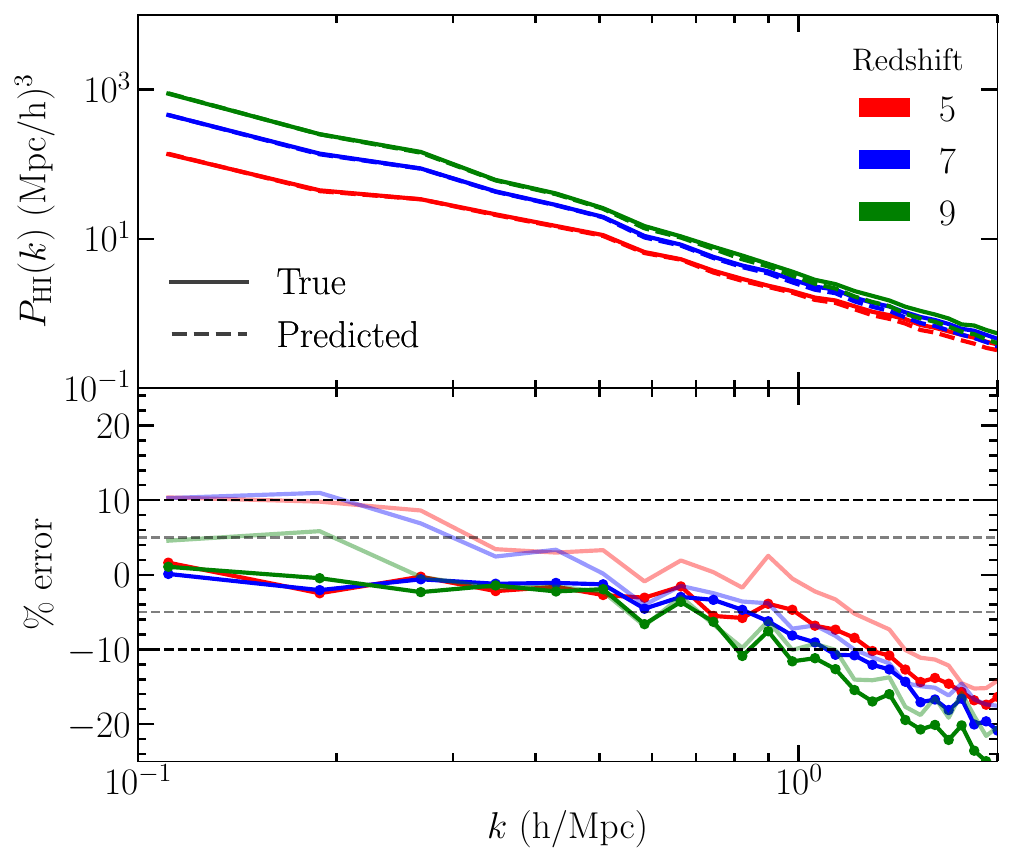}
  \end{subfigure}
  \hspace{0em}
  \begin{subfigure}{0.49\linewidth}
    \includegraphics[width=\linewidth]{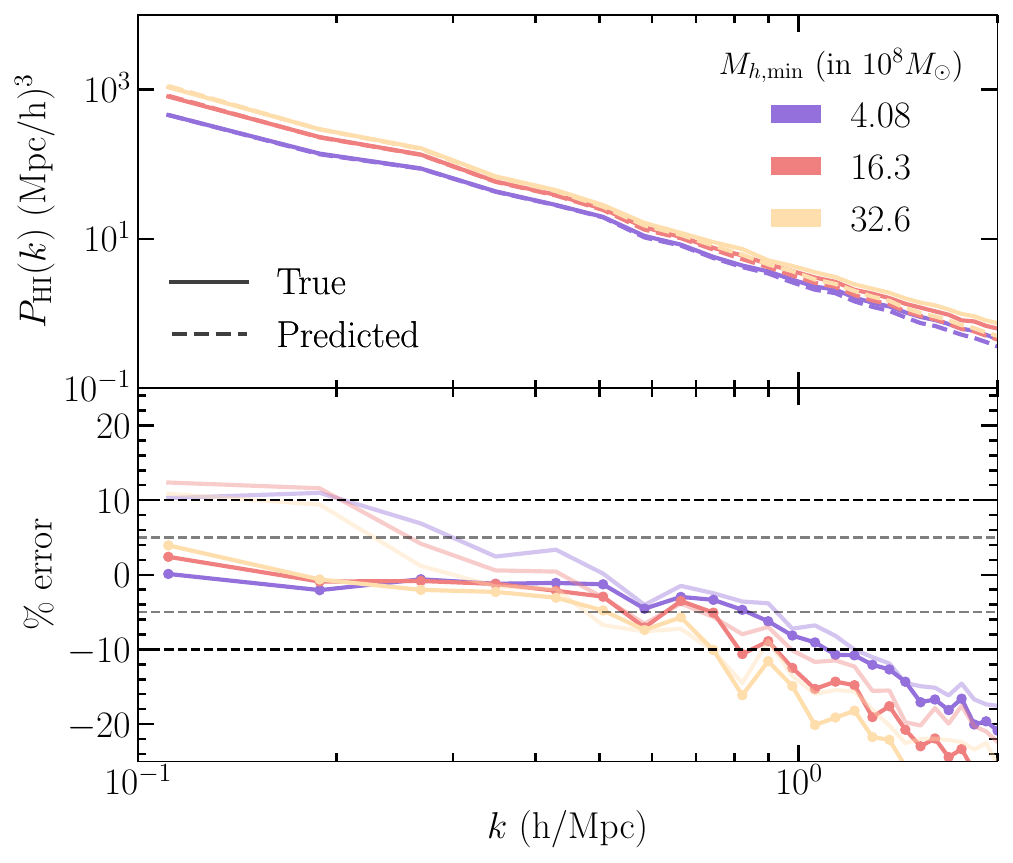}
  \end{subfigure}

  \vspace{-1em}
  
  \caption{\small Neutral hydrogen power spectra $P_{\text{HI}}(k)$ obtained using \textit{binning scheme A} with $(20, 15, 30)$ bins for variations in redshift \textit{(left panel)} and minimum halo mass \textit{(right panel)}. The lower panel shows the relative error with the true $P_{\text{HI}}(k)$ computed from the RB, with the translucent lines showing results from Paper1. Dashed gray and black horizontal lines mark 5\% and 10\% error, respectively. Our method works well for large scales of $k<0.5\ \hMpc$, where the $z$ and $M_{h, \text{min}}$ variations stay within 3\% and 5\% error, respectively. Since we sample \fcoll\ deterministically conditioned on the tidal eigenvalues of the cell, the small-scale power $(k \gtrsim 1\ \hMpc)$ is not recovered as accurately (error $\gtrsim 10\%$). Paper1 results have not been shown in the upper panels of both the plots.}
  \label{z_Mhmin_var}
\end{figure}

We now move to the HI power spectra. The results for redshift $(z)$ and minimum halo mass $(\mhmin)$ variations, with binning scheme A, are shown in figure \ref{z_Mhmin_var}. To demonstrate the improvement, we also include the power spectra errors obtained using the method of Paper1 in the lower panels using translucent lines with the same color corresponding to each case. We observe a remarkable sub-3\% accuracy of the HI power at large scales below $k=0.5\ \hMpc$ in the $z$ variations. Over the same $k$ range, the $M_{h, \text{min}}$ variations show a slightly larger but still very good agreement within 5\%. In all the variations, we see that the agreement degrades quickly at larger $k$ values and becomes $\gtrsim10\%$ beyond $k=1\ \hMpc$. This is expected based on our argument from above --- the small-scale features of the \fcoll\ and HI maps cannot be captured by a deterministic sampling such as ours and the HI power spectra are bound to show a large error at small scales or high $k$. 

\begin{figure}[h]
  \centering
  \begin{subfigure}{0.49\linewidth}
    \centering
    \includegraphics[width=\linewidth]{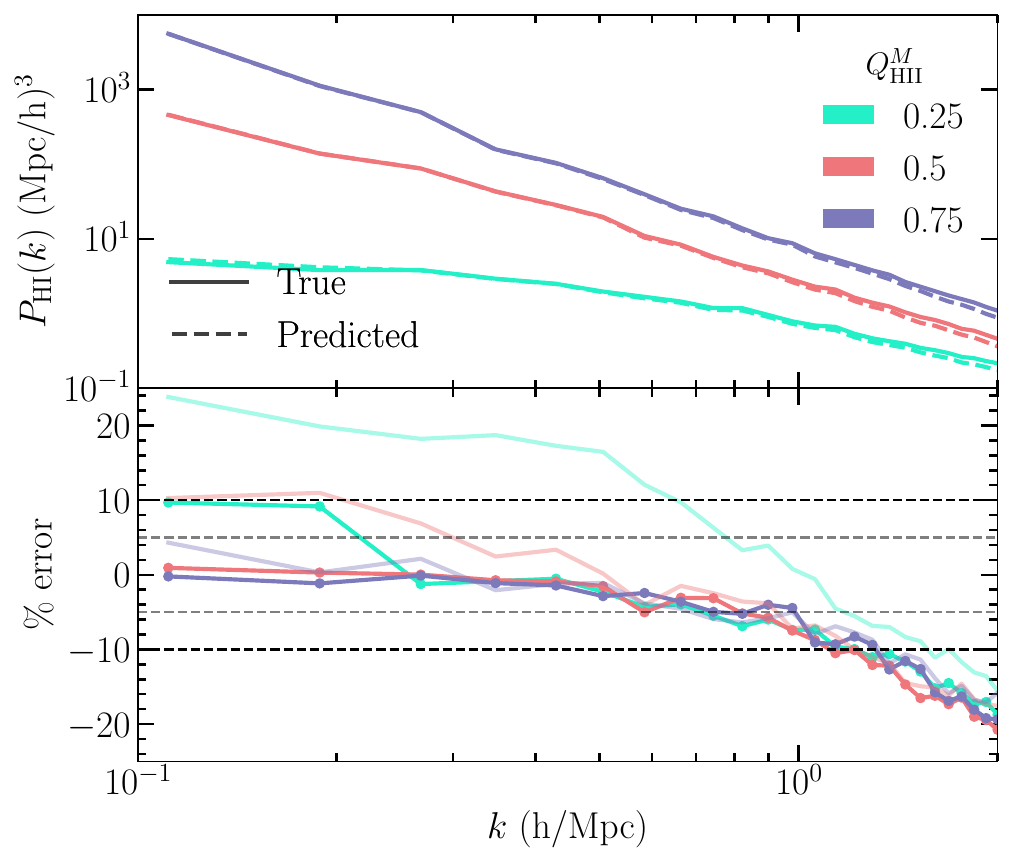}
  \end{subfigure}
  \hspace{0em}
  \begin{subfigure}{0.49\linewidth}
    \includegraphics[width=\linewidth]{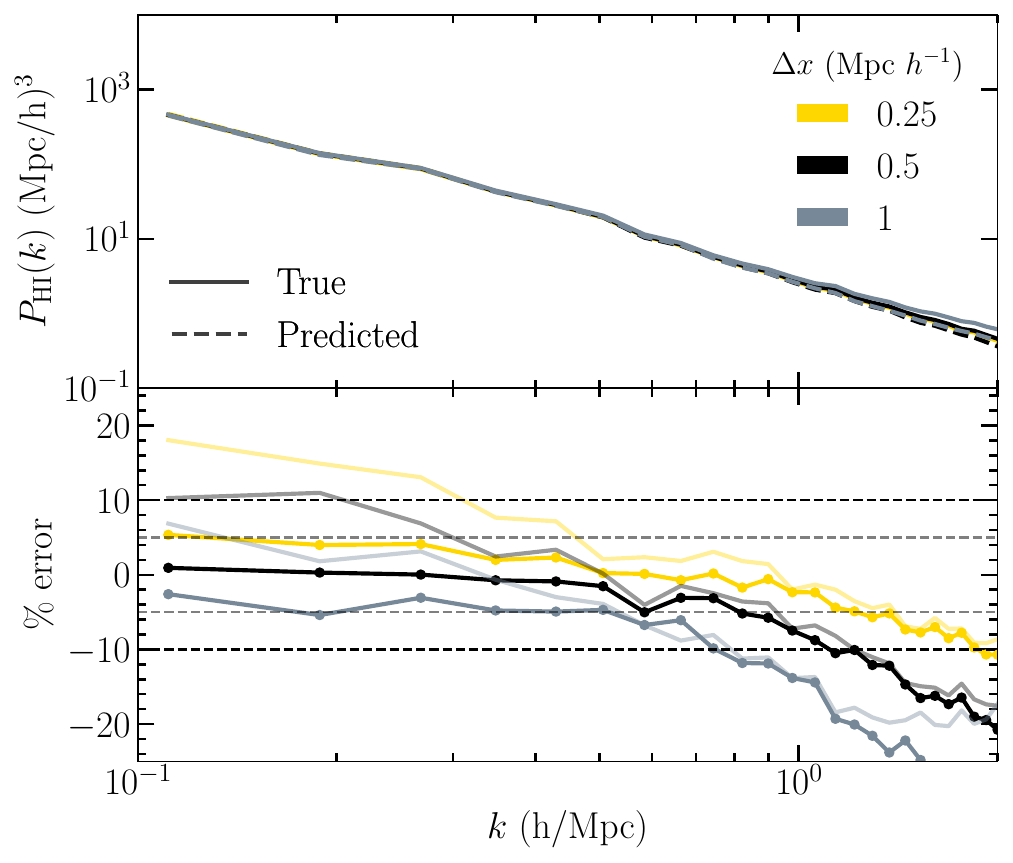}
  \end{subfigure}

  \vspace{-1em}
  
  \caption{\small Neutral hydrogen power spectra $P_{\text{HI}}(k)$ obtained using \textit{binning scheme B} with $(25, 15, 20)$ bins for variations in ionized fraction \textit{(left panel)} and grid size \textit{(right panel)}. The lower panel shows the relative error with the true $P_{\text{HI}}(k)$ computed from the RB, with the translucent lines showing results from Paper1. Dashed gray and black horizontal lines mark 5\% and 10\% error, respectively. The $\qmhii = 0.75$ case shows a great agreement with $<3\%$ errors upto $k=0.5\ \hMpc$, while the $\qmhii = 0.25$ case has a relatively larger error of $\sim 9\%$ for $k\leq 0.2\ \hMpc$ that improves to $<3\%$ upto $k \leq 0.5\ \hMpc$. The $\Delta x = 1\ \Mpch$ and $\Delta x = 0.25\ \Mpch$ cases show errors of $-5\%$ and $+5\%$, respectively at large scales upto $k = 0.5\ \hMpc$.  }
  \label{qmhii_dx_var}
\end{figure}

Let us now focus on the results for variations in ionized fraction (\qmhii) and grid size $(\Delta x)$, obtained using binning scheme B and shown in figure \ref{qmhii_dx_var}. As in figure 2, Paper1 results have been included in the error panels using translucent lines. In the \qmhii\ variations, the 0.75 case does really well with almost sub-2\% errors across most of the $k$ range below 0.5\ \hMpc. On the other hand, the $\qmhii = 0.25$ variation has an error of $\sim 9\%$ at the largest scales $(k \leq 0.2\ \hMpc)$ and subsequently drops down to sub-3\% levels for $k \leq 0.5\ \hMpc$. The $\Delta x$ variations have slightly larger errors but still remain within 5\% in magnitude, with the 0.25\ \Mpch variation being $+5\%$ and the 1\ \Mpch one being $-5\%$ over the same $k<0.5\ \hMpc$ range. Both the \qmhii\ and $\Delta x$ variations show the expected increase of error beyond $\sim 10\%$ at sufficiently high $k$. The exact scale at which this happens is quite different between the various cases of the $\Delta x$ variation, simply because of their different Nyquist frequencies and the sampling introducing discrepancies primarily at the scale of a few, neighbouring cells.

\section{Discussion and Applications}
\label{sec:disc}

Our previous work in Paper1 had the same goal of modeling the HI power spectra accurately across variations in certain physical and simulation parameters, while using a similar approach of combining two low dynamic range boxes (`SB and LB'). The only difference was the physical variable conditioning the collapse fraction \fcoll\, which was taken to be the dark matter density contrast $\delta$. Here, we extend this approach to include more local information about the cosmological environment that can potentially affect the distribution of halos. This is done by conditioning the \fcoll\ values on linear combinations (given in equations \ref{lin_combs_eqn1}--\ref{lin_combs_eqn3}) of the three eigenvalues of the tidal tensor as defined in equation \ref{tidal_tensor_eqn}. 

We find significant improvements in the HI power spectra relative to the results of the previous study described above. Table \ref{improvement_summ_table} summarizes the improvements in the relative errors of $P_\text{HI}(k)$ over two separate $k$ ranges.

\begin{table}[ht]
\centering
\caption{Error improvement in the HI power spectra (previous work $\bm{\rightarrow}$ current work).}
\label{improvement_summ_table}
\begin{tabular}{|c|c|c|}
\hline
\textbf{Case} & \boldmath{$k \leq 0.2\ \hMpc$} & \boldmath{$0.2 < k \leq 0.5\ \hMpc$} \\
\hline
Fiducial      & $\sim 10\% \bm{\rightarrow} \lesssim 2\%$ & $\sim 5\% \bm{\rightarrow} \sim 1\%$ \\
$z=5$        & $\sim 10\% \bm{\rightarrow} \lesssim 2\%$ & $\sim 5\% \bm{\rightarrow} \lesssim 2\%$ \\
$z=9$        & $\sim 5\% \bm{\rightarrow} \sim 1\%$      & similar \\
$\mhmin=16.3 \times 10^8\ \Mh$    & $\sim 12\% \bm{\rightarrow} <4\%$         & marginally better \\
$\mhmin=32.6 \times 10^8\ \Mh$    & $\sim 12\% \bm{\rightarrow} <4\%$         & marginally better \\
$\qmhii=0.25$         & $\gtrsim 20\% \bm{\rightarrow} \sim 9\%$  & $\sim 18\% \bm{\rightarrow} <3\%$ \\
$\qmhii=0.75$         & marginally better                  & similar \\
$\Delta x = 0.25\ \Mpch$ & $\sim 15\% \bm{\rightarrow} \lesssim 5\%$ & $\sim 8\% \bm{\rightarrow} \sim 2\%$ \\
$\Delta x = 1\ \Mpch$    & similar                           & similar \\
\hline
\end{tabular}
\end{table}

As mentioned in section \ref{sec:intro}, the authors of \cite{barsode_24} also use the eigenvalues of the tidal tensor to inform their cell-wise prediction of collapse fraction, although using a different `matching method' that makes use of a low-volume simulation box corresponding to the one for which the prediction has to be made. It is important to note that we can only compare our results for the HI power spectra with theirs in an approximate sense, since they do not use the same values for $\Delta x$ and \mhmin\ as we do. The closest parameter combination that we can compare our fiducial case with is $z=7, \mhmin=8.15 \times 10^8\ \Mh, \Delta x = 0.62\ \Mpch$. We find that our method produces a smaller magnitude of error ($\sim 1\%$) at larger scales ($k\lesssim 0.6\ \hMpc$), while the matching method works better at smaller scales ($\sim 5\%$ for $k \gtrsim 0.9\ \hMpc$). Similarly, the \qmhii variations of 0.25 and 0.75 perform better with our method at low $k$ values below roughly $0.6\ \hMpc$ and achieve $\sim 2\%$ accuracy, but for larger $k$ values $(\gtrsim 1\ \hMpc)$ their error increases drastically while the matching method persists at around $\lesssim 5\%$ for $\qmhii=0.75$ and at around $8\%$ for $\qmhii=0.25$. Our results for the $\Delta x = 0.25\ \Mpch$ case are better with $<5\%$ errors over a wide $k$ range where their $\Delta x = 0.31\ \Mpch$ case consistently shows $>5\%$ errors. However, their $\Delta x = 1.25\ \Mpch$ case outperforms our $\Delta x = 1\ \hMpc$ case by having sub-5\% errors down to $k=2\ \hMpc$. The method in \cite{barsode_24} transfers the complete halo catalog from a high-resolution cell in their small box to its tidal-environment-matched counterpart in the low-resolution box. This process preserves the sub-grid variance in the collapse fraction, which is averaged out in our conditional mean approach, thus allowing their method to achieve better accuracy at high $k$. Making comparisons with the minimum halo mass and redshift variations is simply not possible due to substantially different parameter combinations than used in this work. 

It is worth emphasizing the simplicity of our method, where we do not train any machine learning algorithm but just compute conditional means of \fcoll\ over an optimized binning. Once the SB and LB simulations are available, the entire process of making the prediction from computing the conditional means to sampling them takes no more than 5 minutes. This makes it a highly efficient method for RAM-limited users that can run boxes like LB or SB with a lower resource requirement than the full RB. The following future directions can be explored further using our method ---

\begin{itemize}
    \item Using the fast and fairly accurate predictions of the $\fcoll(\mathbf{x})$ field at, say $z=5$ and $z=7$, one can think of an interpolation scheme to approximate the \fcoll\ field at an intermediate redshift, say $z=6$. If this can be done while achieving reasonable errors for the corresponding HI power spectrum, relying on the fact that the errors in the $z=5$ and $z=7$ cases are extremely low, then it eliminates the need for a separately optimized \fcoll\ conditional mean matrix at $z=6$. 
    \item Implementing the method on a larger box with a different seed: if our method is robust to cosmic variance, it should be directly applicable to boxes run using different initial conditions than our RB. It will be interesting to apply our technique to the $200\ \Mpch$ boxes run using $2048^3$ particles as part of the \texttt{Sahyadri} simulation suite (Dhawalikar et al., in prep.).
    \item The same technique can be applied to LB simulations with different cosmological parameters (cf. the \texttt{Sahyadri} suite), for which the SB simulations would need to be performed separately. Interpolations similar to those discussed above for multiple redshifts can then be envisaged, as a stepping stone to building an efficient emulator of $\fcoll(\mathbf{x})$.
\end{itemize}

\section{Conclusion}
\label{sec:conc}

In this work, we have presented an efficient and accurate method for predicting the neutral hydrogen (HI) density fields and their corresponding power spectra at large scales during the Epoch of Reionization (EoR). Modeling the EoR is challenging due to the computational complexity of full radiative transfer simulations \cite{gnedin_00, ciardi_03, mellema_06, trac&cen_07, gnedin_14, rosdahl_18, lewis_22} and inaccuracies of semi-analytical models used to prescribe the collapse fraction \cite{PS_74, BCEK_91, ST_99, ST_02, mcquinn_07, doussot_22}. Building upon earlier approaches that rely solely on the dark matter overdensity for conditioning the collapse fraction distribution \cite{ahn_12, iliev_14, pundir_25}, we show that incorporating information from the eigenvalues of the tidal tensor significantly improves the accuracy of the HI power spectrum at large scales. 

We employed a simple deterministic sampling method based on the mean collapse fraction given the eigenvalues $\avg{\fcoll|\bm{\ell}}$ derived from a suite of smaller, high-resolution N-body simulations, with the prediction itself made using the eigenvalues from a low-resolution, large-volume simulation. We optimize the binning of the eigenvalues in a way that produces the least error for the HI power spectra, $P_\text{HI}(k)$. The results demonstrate a significant improvement in the accuracy of $P_\text{HI}(k)$ at large scales $(k \leq 0.5\ \hMpc)$, with errors typically being around 2\%--5\% across a wide range of physical and simulation parameters, including redshift, ionized fraction, grid resolution, and minimum halo mass (figures \ref{z_Mhmin_var} and \ref{qmhii_dx_var}). This is a marked improvement over our previous work, which relied solely on the dark matter density contrast to condition the \fcoll\ and achieved $\sim 10\%$ error in the large-scale HI power spectrum (table \ref{improvement_summ_table}). The limitation of our deterministic method is that it inevitably smooths out small-scale fluctuations, leading to poor accuracy at higher wavenumbers. 

The key advantages of our approach are its simplicity and computational efficiency. Once the conditional means are tabulated from high-resolution small-box simulations, predictions for large volumes can be generated within a couple of minutes, without needing sophisticated machine learning algorithms or simultaneous high- and low-resolution runs. This makes the method well-suited for fast parameter-space exploration and for producing HI density field realizations in RAM-limited settings. Future applications of this work could involve extending it to larger simulation volumes to test for robustness against cosmic variance, and developing interpolation schemes for \fcoll\ predictions across different redshifts and cosmologies to eliminate the need for separate conditional mean evaluations. 

\acknowledgments

The research of AP is supported by the Associateship Scheme of ICTP, Trieste.  We gratefully acknowledge computing facilities at NCRA for running the GADGET-2 simulations. The resources provided by the PARAM Brahma facility at IISER Pune which is a part of the National Supercomputing Mission (NSM) of the Government of India are also gratefully acknowledged. 

\section*{Data availability}

The tidally binned $\avg{\fcoll|\bm{\ell}}$ matrix is available upon reasonable request to the authors, and will eventually be incorporated into the publicly available \textsc{script} code.

\bibliography{references}

\appendix

\section{Robustness of Binning}\label{app:robustness}

In section \ref{sec:results}, we identified our two main binning schemes (called A and B) by recording the number of logarithmic bins along each of the variables $(\ell_1, \ell_2, \ell_3)$ defined in equations \ref{lin_combs_eqn1}--\ref{lin_combs_eqn3} that produce the lowest error in the HI power spectrum for a range of respective parameter variations. We now show how our results vary if we instead use the binning schemes in the neighborhood of A and B.

\begin{figure}[h]
  \centering
  \begin{subfigure}{0.49\linewidth}
    \centering
    \includegraphics[width=\linewidth]{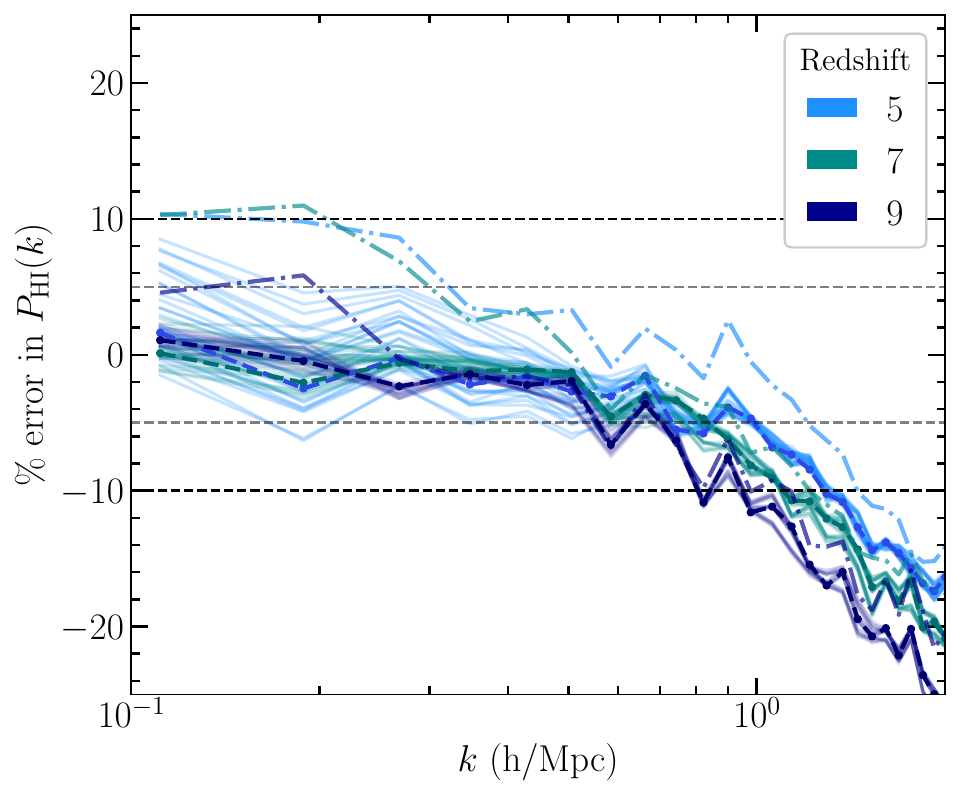}
  \end{subfigure}
  \hspace{0em}
  \begin{subfigure}{0.49\linewidth}
    \includegraphics[width=\linewidth]{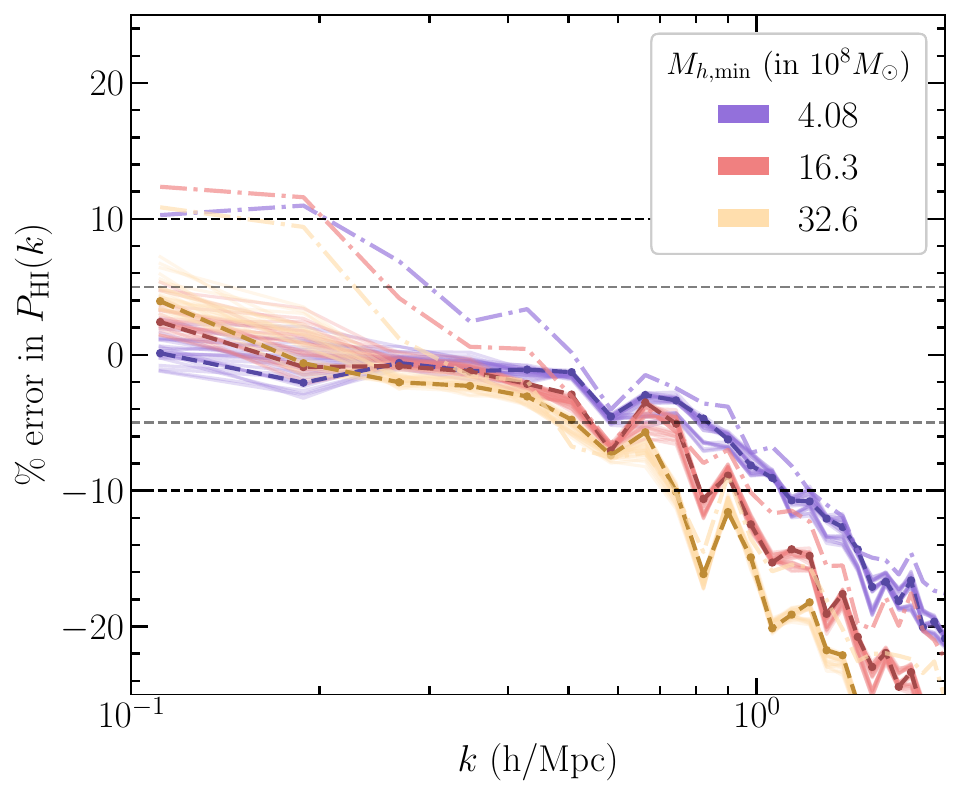}
  \end{subfigure}

  \vspace{-1em}
  
  \caption{\small Errors in HI power spectra $P_{\text{HI}}(k)$ obtained using 27 different binning schemes in the neighborhood of and including scheme A, presented for the variations in redshift \textit{(left panel)} and minimum halo mass \textit{(right panel)}. Scheme A errors are in darker, dashed lines and the solid translucent lines around them show errors using the nearby schemes. Paper1 results have been shown in dashdot lines for reference. }
  \label{sen_anal_z_Mhmin}
\end{figure}

\begin{figure}[h]
  \centering
  \begin{subfigure}{0.49\linewidth}
    \centering
    \includegraphics[width=\linewidth]{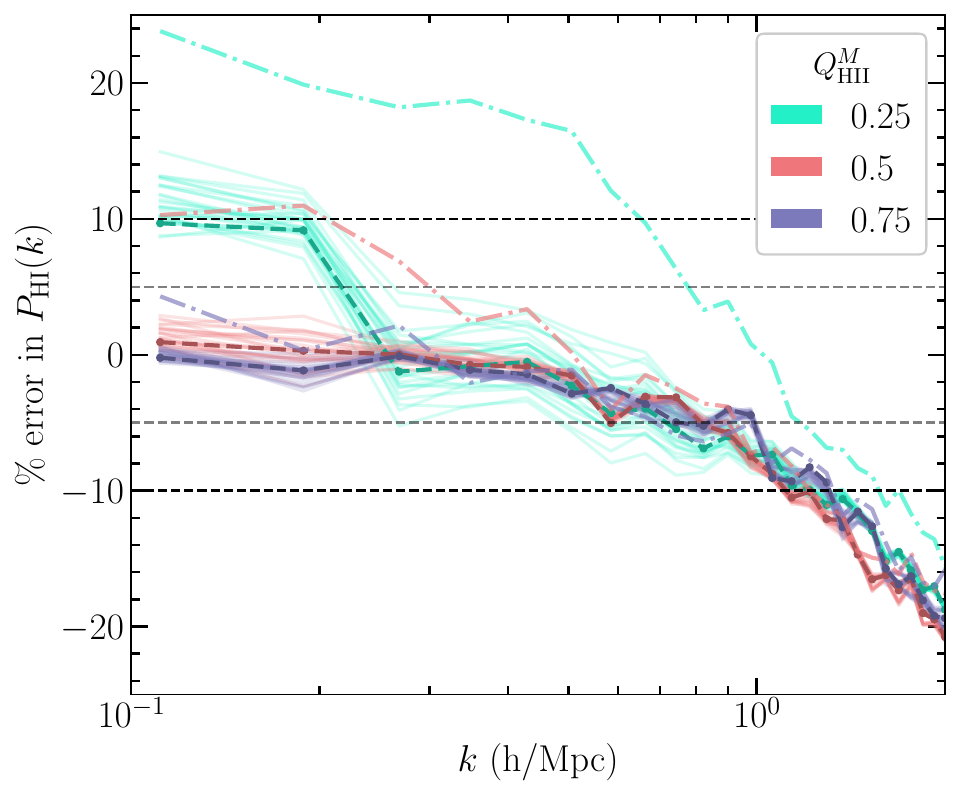}
  \end{subfigure}
  \hspace{0em}
  \begin{subfigure}{0.49\linewidth}
    \includegraphics[width=\linewidth]{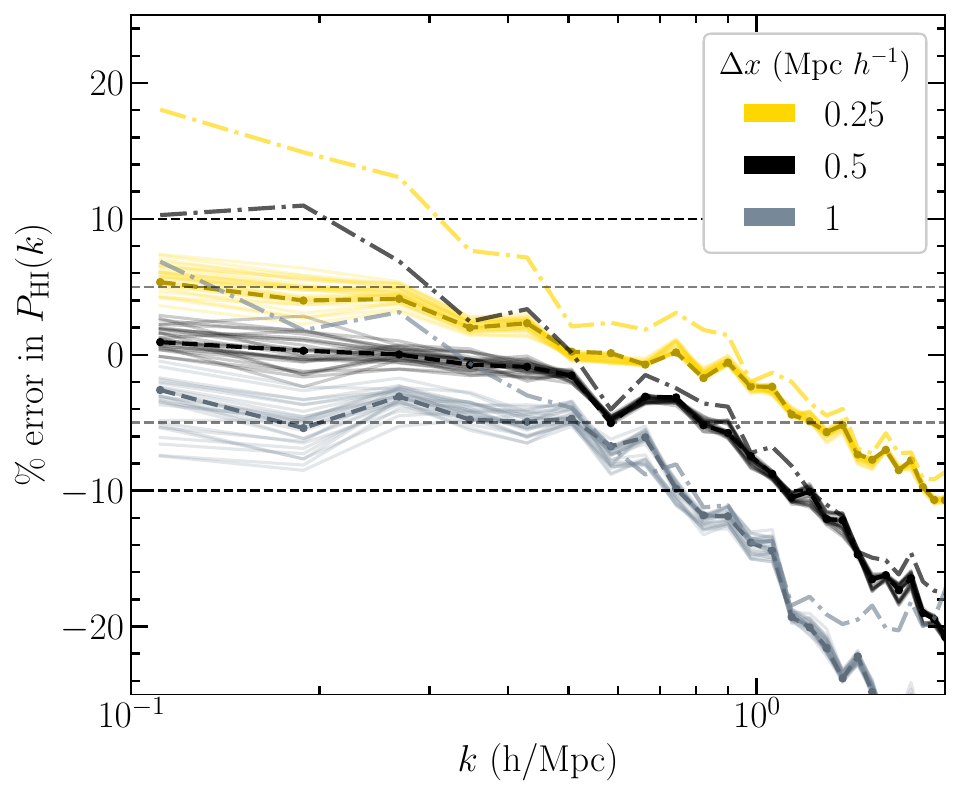}
  \end{subfigure}

  \vspace{-1em}
  
  \caption{\small Errors in HI power spectra $P_{\text{HI}}(k)$ obtained using 27 different binning schemes in the neighborhood of and including scheme B, presented for the variations in ionization fraction \textit{(left panel)} and grid size \textit{(right panel)}. Scheme B errors are in darker, dashed lines and the solid translucent lines around them show errors using the nearby schemes. Paper1 results have been shown in dashdot lines for reference.}
  \label{sen_anal_qmhii_dx}
\end{figure}

Scheme A contains $(20, 15, 30)$ bins and so we consider all schemes upto the third nearest neighbor, i.e. by making all combinations varying the number of bins independently along each direction by $+5$ and $-5$. This gives 3 choices (including the fiducial) for each direction, leading to a total of 27 schemes. We similarly obtain 27 schemes around and including scheme B which has $(25, 15, 20)$ bins, and overlay the HI power spectra errors obtained from all the neighboring schemes with respect to the reference RB power spectra. This is presented in figure \ref{sen_anal_z_Mhmin} (figure \ref{sen_anal_qmhii_dx}) for perturbations around scheme A (scheme B), corresponding to variations in redshift and minimum halo mass (ionization fraction and grid size). 

The optimized schemes A and B themselves are shown in the figures using dashed curves in a slightly darker shade of the respective colors, and the results from Paper1 have also been overlaid in dashdot curves. The robustness of the method is most clearly evident in the cases of $\qmhii=0.25$ and $\Delta x=0.25$ (figure \ref{sen_anal_qmhii_dx}), where most of the neighboring schemes have their power spectrum errors lying in a range of 3--4\% around the scheme B errors, which are $\sim9\%$ ($\sim 5$\%) as opposed to the $\sim 20\%$ ($\sim 16\%$) from Paper1 for $\qmhii=0.25$ ($\Delta x=0.25$). Similarly, the nearby schemes in the \mhmin\ variations (excluding the fiducial) in figure \ref{sen_anal_z_Mhmin} lie in a range of 3--4\% around the scheme A error which is 1--2\% in contrast to the 10--12\% from Paper1. The fiducial case also shows similar numbers to the \mhmin\ variations for both, neighbors of scheme A and B. Finally, the $z=5$ ($z=9$) case has a spread of errors that is $\sim 5\%$ ($\sim 2\%$) wide around the optimal results of $\sim1\%$, which is a substantial improvement from the $\sim 10\%$ ($\sim 5\%$) value from Paper1.

In conclusion, while the neighboring schemes do show some variance around the errors using the optimized scheme, the spread is small compared to the difference with the Paper1 results for all parameter variations and scales where scheme A or B perform better than Paper1.

\end{document}